\documentclass[showpacs,aps,prd,eqsecnum,floats,preprint]{revtex4}

\newif\ifpdf\ifx\pdfoutput\undefined\pdffalse\else\pdfoutput=1\pdftrue\fi

\usepackage{graphicx,bm,dcolumn}
\graphicspath{{./graphics/}}

\ifpdf\usepackage{hyperref}\else\fi
\ifpdf\DeclareGraphicsExtensions{.pdf,.jpg,.png}\else\fi
\def\fcut{f_{\mathrm{cut-off}}}
\def\msun{\mathrm{M}_\odot}
\def\Hz{\mathrm{Hz}}
\def\smin{S_{\mathrm{min}}}
\def\sobs{S_{\mathrm{obs}}}
\def\sec{\mathrm{sec}}

\def\nop{{\mathcal{N}}_{\mathrm{op}}}
\def\nflop{{\mathcal{N}}_{\mathrm{flop}}}

\def\fsamp{f_{\mathrm{samp}}}
\def\htilde{\tilde{h}(f)}
\def\amb{{\mathcal{H}}}
\def\ntemp{{\mathcal{N}}_{\mathrm{t}}}
\def\be{\begin{equation}}
\def\ee{\end{equation}}
\def\dth{\Delta \theta}
\def\lsim{\mathrel{\rlap{\lower4pt\hbox{\hskip1pt$\sim$}}\raise1pt\hbox{$<$}}}                
\def\gsim{\mathrel{\rlap{\lower4pt\hbox{\hskip1pt$\sim$}}\raise1pt\hbox{$>$}}}                
\def\tpad{T_{\mathrm{pad}}}
\def\npad{N_{\mathrm{pad}}}

\begin{document}

\preprint{IUCAA 02/2003} 

\title{A faster implementation of the hierarchical search algorithm for detection of 
gravitational waves from inspiraling compact binaries
}

\author{
Anand S. Sengupta \email{anandss@iucaa.ernet.in}
}\affiliation{Inter-University Centre for Astronomy and Astrophysics,\\ Post 
Bag 4, Ganeshkhind, Pune 411007, India}

\author{ Sanjeev Dhurandhar \email{sanjeev@iucaa.ernet.in}
}
\affiliation{Inter-University Centre for Astronomy and Astrophysics,\\ Post 
Bag 4, Ganeshkhind, Pune 411007, India}

\author{Albert Lazzarini \email{lazz@ligo.caltech.edu}
}
\affiliation{LIGO Laboratory, California Institute of Technology,\\
MS 18-34, Pasadena, CA 91125, United States}

\begin{abstract}

The first scientific runs of kilometer scale laser interferometric detectors like LIGO are underway. 
Data from these detectors will be used to look for signatures of gravitational waves (GW) from 
astrophysical objects like inspiraling neutron star/blackhole binaries using  
matched filtering. The computational resources required for  online flat-search implementation of the 
matched filtering are large if searches are carried out for small total mass. 
Flat search is implemented by constructing a single discrete grid of 
densely populated template waveforms
spanning the dynamical parameters - masses, spins - which are correlated with the interferometer data.
The correlations over the kinematical parameters can be maximized apriori 
without constructing a template bank
over them.
Mohanty and Dhurandhar (1996) showed that a significant reduction in computational resources can be 
accomplished by using a hierarchy 
of such template banks where candidate events triggered by a sparsely populated  grid is followed up by 
the regular, dense flat search grid. The estimated speed up in this method was  a factor $\sim 25$ over the 
flat search.  
In this paper we report an improved implementation of the hierarchical search, wherein we extend the 
domain of hierarchy to an extra dimension - namely the time of arrival of the signal in the bandwidth of
the interferometer. This is accomplished by lowering the Nyquist sampling rate of the signal in the trigger stage. 
We show that this leads to  further improvement in the efficiency of data analysis and speeds up the online
computation by a factor of $\sim 65 - 70$ over the flat search. We also take into account and discuss issues
related to template placement, trigger thresholds and other peculiar problems that do not arise in earlier 
implementation schemes of the hierarchical search. We present simulation results for 2PN waveforms embedded 
in the noise expected for initial LIGO detectors.

\end{abstract}
\pacs{04.80.Nn, 07.05.Kf, 95.55.Ym, 97.80.-d}


\maketitle
\section{Introduction}
\label{sec:intro}
Pulsar timing experiments by Hulse and Taylor \cite{HT1,HT2} led to accurate measurement of periastron time 
shifts in
the PSR1913+16 binary pulsar system. These matched to the predictions from the theory of general relativity 
for energy losses due to gravitational waves (GW) to better than $0.5 \%$ accuracy. This was the first strong
indirect evidence for the existence of GW. Over the last decade or so, a host of laser interferometric
detectors like the two LIGO detectors \cite{LIGO} in USA, GEO600 in Germany \cite{GEO600}, the TAMA 300 in 
Japan \cite{TAMA300} and VIRGO in Italy-France \cite{VIRGO} are
being built to {\em catch} these waves {\em in flesh} and unravel the physics of the universe encoded
in them. As of present, several of these interferometers  are in advanced stage of completion. 
The TAMA 300 has achieved significant observation time with more than 1000 hours of observation whereas the 
LIGO had its first successful scientific run in August, 2002. Thus data from these detectors 
promise much interesting and new astronomical research over the next several years.

The most well studied sources of GW important to detection are compact astrophysical objects moving at 
relativistic speeds, like merging blackholes, spinning neutron star systems, inspiraling neutron star binaries 
etc.
\cite{Thorne,sources}.  
A gravitational wave burst with a typical GW strain of $h \sim 10^{-21}$  is expected to 
carry \cite{weiss} about $80 \times 10^{-6}\ {\mathrm{watts/meter^2}} $ 
past the modern detectors over a duration of $\sim 10$ millisec. However, GW interact so 
weakly with matter that even this huge flux of energy (by electromagnetic  standards) will produce only a minute 
measurable signal of the above order in the arm lengths of the detectors.
Detection of such weak signals requires special signal 
processing methods to extract the signature of these 
signals buried in the noisy output of these detectors.

Accurate waveform modeling of gravitational waves from the inspiral phase of compact binaries 
\cite{pn2} 
makes it possible to look for their signatures using the technique of matched filtering and are 
considered to be the likely sources to be detected by the first generation of detectors. In the
last few minutes of their inspiral, the frequency of the emitted GW will lie in the sensitive bandwidth
of these detectors. Matched filtering involves cross correlating the detector output with a bank of 
template waveforms each having a different set of parameters. This is necessary since only 
the {\em form} of the expected signal is known and not its exact parameters {\em a priori}.  
Together, these templates span the whole range of  parameters  as determined from astrophysical 
considerations. This procedure is computationally less efficient if implemented in a {\em{flat}}
method which uses a very dense grid of templates to span the parameter space. The method is very
well described in earlier literature - see \cite{lal,grasp,sd1,sd2} for example. A more efficient 
implementation of the flat search is desirable for several
reasons. Eventually the interferometers will run continuously and  real time analysis must be 
carried out in order to keep up with the data streams. This means that the time taken to analyze a data 
segment of duration $T$ seconds, must be $\leq \ T$ 
seconds in real time. By the same token, the computational efficiency of real time algorithms will 
limit the volume of the parameter space which one can search for GW signals in interferometer data. 
The desirability for the improved computational efficiency promised by
the hierarchical algorithm presented here is twofold: (i) to date, there
has been
little or no optimization of the implementation of parallelized
matched
filtering algorithms and their performance will not scale with larger
numbers of
processors \cite{amdahl}; (ii) recent
developments in the methodologies to handle more massive compact binary
systems and to include non aligned spin have shown the need
for a greater number of templates in a higher dimensional search space
than has been applied in the first searches for lighter mass systems \cite{buonnano, owen3}.

A faster implementation of the matched filtering procedure was proposed by Mohanty and Dhurandhar
\cite{monty1,monty2} which could reduce the computational cost of the search algorithm by a factor of 25
to 30 without sacrificing the efficiency of the flat-search method. It essentially involved using 
hierarchically staged analyses on two 
template grids instead of one dense grid. The first layer was that of trigger stage templates which 
were placed more sparsely than the second stage templates. Any {\em{crossing}} over the trigger threshold 
would be followed up {\em{locally}} by the fine bank. This trigger threshold must be carefully chosen such
that not only could  false alarms be kept in  control,  but also the detection probability of a
genuine signal will be higher than a prescribed minimum (usually $\sim 95 \%$). The need for further
improving upon the efficiency of implementing the matched filtering method is more pressing at present than
ever - mainly because of our experience with real interferometer data over the recent past. Some of the benign
assumptions which are used in prescribing the canonical search actually do not hold true in real time data
- for example the stationarity and Gaussianity of interferometer noise is a key assumption to almost all the previous
work. However, drifts in the noise power spectra and {\em{tail}} features therein severely undermine the
efficiency of the matched filtering algorithm \cite{tanaka}. Thus hidden costs are incurred in carrying out 
even a flat
search where $\chi^2$ vetoing \cite{allen} is now prescribed to curb high unprecedented false alarm rates. 
The TAMA data analysis team \cite{tama} had to put such vetoing even in the trigger stage of 
the hierarchical search. Such vetoing techniques add to the total cost of the online computational budget. 
An alternative implementation to the matched filtering has been  proposed recently  
using the Fast Chirp Transform (FCT) method \cite{fct}. 
The complexity of this algorithm is comparable to that of the
conventional matched filtering technique. It may become simpler under
certain circumstances if it is possible to transpose the order in which
the  associated 2-D FFT is calculated. To date, however, this has not
been explored further.

In this paper, we propose a faster implementation of the canonical hierarchical search method. 
Because most of the power in the chirp signal is at low frequencies - 
the power in the signal falls off as $f^{-7/3}$ where $f$ is the frequency - 
one can truncate the Fourier transformed data at a relatively lower
frequency (we choose 256 Hz) and nonetheless retain sufficient signal
power $(\sim 92\%)$ \cite{ourpaper}. Then the key idea is to sample the data at a lower rate
(512 Hz) in the trigger stage which  gives us smaller FFT data sets to work
upon. 
Lowering the first stage sampling rate  also leads to a changed filter placement scenario where the absence of higher frequency
modes of the GW chirp causes the ambiguity function to fall off more gradually and thus individual trigger 
templates can cover a bigger volume of the parameter space. This also leads to further computational advantage since the
number of trigger templates is reduced. However, the lowered trigger thresholds and reduced signal to 
noise ratio due to this method will tend to increase the false alarm rate. We discuss these
issues and supplement our solution with simulation results. In an earlier paper \cite{ourpaper} we had outlined  
some of
the above ideas. However, the scope of that paper did not include such complications
as rotation of the parameter space, innermost stable circular orbit (ISCO) frequency cutoffs etc. Here, we take into 
account all these realistic
scenarios to propose an extended hierarchical search (EHS) algorithm which can reduce the computational
cost by a factor of $\geq$ 65 over a flat search. The family of restricted, spin-less
2PN family of chirp waveforms is used for our computation. We also use the target noise power spectral
density of initial 4 km LIGO interferometers in our simulations \cite{noise}.

\section{The flat search}

In this section, we at first gloss over the necessary background and notation in order to make the 
paper self contained. We then estimate the online computational cost of the flat search.
This will prepare us for the EHS and comparing the cost advantages thereof. Generic time domain 
functions $h(t)$ will be denoted by $\htilde$ in the Fourier domain, where
\begin{equation}
\htilde = \int_{-\infty}^{\infty}dt\ h(t) e^{2\pi ift}.
\label{eq:fftdef}
\end{equation}
Engineering and much of LIGO analysis software uses the opposite sign for the exponent in (\ref{eq:fftdef}).
We however maintain this notation for consistency with the published literature \cite{tanaka, monty1}.

\subsection{Matched filtering}
The stationary phase approximated Fourier transform $\htilde$ of the spin-less, restricted second 
post-Newtonian binary chirp is given up-to a constant factor $\mathcal{N}$ by
\be
\htilde = {\mathcal{N}} f^{-7/6} \exp i \left [  
                           -\frac{\pi}{4} - \Phi_a + \Psi(f;M,\eta,t_a) \right ],
\label{eq:htilde}			   
\ee
where $M = m_1 + m_2$ is the total mass of the binary system, $m_1$ and $m_2$ being the 
individual masses of the stars; $\eta$ is the ratio of the reduced mass to the total 
mass; $\Phi_a$ and $t_a$ are respectively the arrival phase and the time when the frequency 
$f$ attains the fiducial value $f_a$. The factor $\mathcal{N}$ depends on the masses, $f_a$ and the 
distance to the binaries. The function $\Psi(f;M,\eta,t_a)$ describes the phase evolution of the 
inspiral waveform and is given by
\begin{eqnarray}
\Psi(f;M,\eta, t_a) &=& 2\pi ft_a + \frac{3}{128 \eta}
           \left [
            (\pi M f)^{-5/3} + \left ( \frac{3715}{756} + \frac{55}{9}\eta \right )(\pi M f)^{-1}
            \right . \nonumber \\
           && \left . - 16\pi(\pi M f)^{-2/3}  
               + \left ( \frac{15293365}{508032} + \frac{27145}{504} \eta
              + \frac{3085}{72} \eta^2 \right )(\pi M f)^{-1/3} 
	  \right ].
\label{eq:phase}
\end{eqnarray}

It is important to distinguish between the nature of parameters that appear in the above {\em form} of
the GW chirp. The two mass parameters $\vec \mu \equiv \{ M,\eta \}$ are the dynamical parameters and 
determine the {\em
shape} of the chirp. Exclusive to this set are the `offset' parameters $\vec \lambda
\equiv \{ t_a,\Phi_a \}$ which determine the duration of the chirp or the end-points in the time series. The latter can be quickly estimated in
the matched-filtering paradigm without having to construct template banks over them. For example,  
spectral correlators using FFTs allows us to estimate correlations at all time lags and thus estimate 
$t_a$. Similarly, using the quadrature formalism to construct template banks, the initial phase can be 
estimated analytically. However, the dynamical parameters need to be tackled by building a grid of 
templates which span them. In future, our parameter-space will refer explicitly to these mass 
parameters unless otherwise mentioned. To ease the computation, a new set of time parameters $\{ \tau_0, \tau_3 \}$
that are functions of the masses
is chosen in which the metric defined over the parameter space is 
approximately constant. They are related to $\{ M, \eta \}$ by
\begin{equation}
\tau_0 = \frac{5}{256 \pi \eta f_a} \left ( \pi M f_a \right )^{-5/3}, \ \ \ \ \ \tau_3 = \frac{1}{8 \eta f_a} 
        \left ( \pi M f_a \right )^{-2/3}.
\end{equation}
In other words, the template bank is set up over $\vec \mu \equiv \{ \tau_0,\tau_3 \}$. The form 
of (\ref{eq:htilde}) allows us to write the explicit quadrature representation of the $i$-th template as
\be
\tilde{h}(f;\vec \mu_i, t_a, \Phi_a) = \tilde{h_c}(f; \vec \mu_i, t_a) \cos \Phi_a  
                                    + \tilde{h_s}(f; \vec \mu_i, t_a) \sin \Phi_a,
\label{eq:quadrature}
\ee
where,
\begin{eqnarray}
\tilde{h_c}(f; \vec \mu_i, t_a) 
         &=& i \tilde{h_s}(f; \vec \mu_i, t_a) \nonumber \\
         &=& {\mathcal{N}} f^{-7/6} \exp i \left [-\frac{\pi}{4} + \Psi(f; \vec \mu_i,t_0) \right ].
\end{eqnarray}

The normalization constant $\mathcal{N}$ in (\ref{eq:quadrature}) is fixed by demanding unit norm of 
both $h_c$ and $h_s$ i.e. $(h_c,h_c) = (h_s, h_s)  = 1$, where
the scalar product $(a,b)$ of two real functions $a(t)$ and $b(t)$ is defined as 
\be
\left ( a, b \right ) = 2 \int_{f_l}^{f_u} df \ \frac{  \tilde{a}(f) \tilde{b}^*(f) \ + \ 
{\mathrm{c.c.}}}{S_n(f)},
\label{eq:scalar}
\ee
where, we use the Hermitian property of Fourier transforms of real functions. $S_n(f)$ is the one
sided power spectral density of the noise of initial LIGO detectors \cite{noise}. Further, 
$f_l \leq f \leq f_u$ is taken to be the effective spectral window for computing the scalar product. 
The lower frequency cut-off $f_l$ is dependent on the sensitivity of the detector to seismic 
vibrations and is taken to be $30\ \Hz$,  whereas the upper frequency cut-off $f_u$ is equal to half the full sampling frequency 
taken to be $2048\ \Hz$ in this paper. In practice, the upper cut-off frequency for the signal is held back from
the Nyquist rate by a factor $\sim 0.8$. Therefore, for normalizing the templates, we choose an upper cut-off
frequency $f_c$ which is the minimum of $800\ \Hz$  and the so called ISCO cut-off frequency.
The latter corresponds to the last stable circular orbit 
at which the inspiral phase is deemed to end and is relevant only for high 
masses when the ISCO frequency is less than $800\ \Hz$. For low masses where the ISCO cut-off frequency is more than 
$800\ \Hz$, the loss in SNR due to this band limiting is $\sim 0.14\%$, which is negligible.

Having thus set up a template bank, the statistic $\rho$ for an output signal $s(t)$
of the interferometer is the maximum of the signal to noise ratio (SNR) over all the templates (labeled 
by $i$) and the time-lags $t_a$. Thus, the statistic $\rho$ is given \footnote{In  much of recent
literature as also in the LIGO Algorithms Library documentation \cite{lal}, the square root in
(\ref{eq:rho}) is not taken. We however, choose to take the square root to facilitate computation of
detection and false alarm probabilities. See \ref{sec:finebank} for a discussion.} by,
\be
\rho = \max_i \left [ \max_{t_a} \sqrt{ (s,h_c)^2 + (s,h_s)^2 } \right ],
\label{eq:rho}
\ee
which is then compared with a pre-determined threshold $\zeta$. 
Any crossing above this 
threshold is recorded as a candidate event in flat search scenarios. Let the SNR be maximized 
by the $\bar{i}$-th template and at the time-lag $\bar{t}_a$. The phase offset of 
the chirp $\Phi_a$ can then be calculated as:
\be
\Phi_a = \tan^{-1} \left [ \frac{ (s,h_s)}{(s,h_c)} \right ]_{\bar{i}, \bar {t}_a}.
\ee
A detailed description of this procedure can be found in \cite{lal,grasp,monty1,sd1,sd2}.

\subsection{Computational cost of flat search}

Since $\vec \mu$ is spanned by a {\em discrete} grid of templates, it might so happen that a signal 
arriving at the interferometer has parameters that do not correspond to a grid point. In this event, the
correlation of such a signal with the nearby templates will fall below the maximum obtainable value. The
degree of discretisation is governed essentially by the rule of thumb - that in the worst case, the loss
in correlation shall not exceed $3\%$ of the maximum. This corresponds to a loss of event rate of
roughly $10\%$ which is an agreed upon number in the community.
Given the range of masses which act as search limits and set the boundary of $\vec \mu$, 
a grid of templates is constructed with the above idea. 
The main computational cost incurred in the flat search method outlined previously is in performing 
the Fast Fourier Transforms (FFTs) and multiplying the elements. If a data segment of duration $T$ , 
sampled over $N$ points is parsed through a flat search, utilizing $\ntemp$ templates, the order of floating point
operations is 
\be
\nop \sim 6 \ntemp N \log_2(N).
\label{oper}
\ee
If these many operations are to be carried out in real time, the computational speed in floating point
operations per second  (Flops) is given by
\be
\nflop = \frac{\nop}{\tpad},
\label{cspd}
\ee
where $\tpad$ is the padding length in the data segment. Computationally it is advantageous to use 
stretches $T$ that are significantly longer than the duration 
$\sim 3\times \Delta T_{\mathrm{chirp}}$, where $\Delta T_{\mathrm{chirp}}$ is the duration of the longest chirp.
However, in order to preclude losing chirps that occur within 
$\Delta T_{\mathrm{chirp}}$ of the ends of the longer data stretch $T$, it is
necessary to overlap adjacent epochs of data by this amount. Thus the effective
computational time available to analyze a data stretch of duration $T$ is $\tpad = T
- \Delta T_{\mathrm{chirp}}$.

\begin{figure}[h]
\centering
\includegraphics[width=0.55\textwidth]{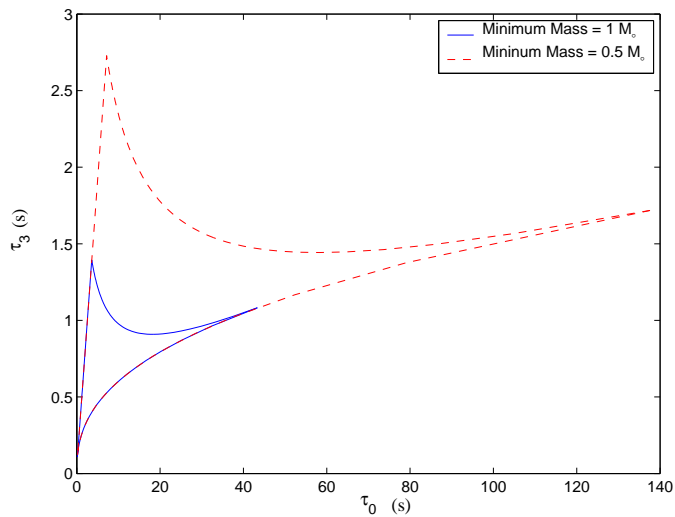}
\caption{Parameter spaces for the individual mass ranges $1-30~ \msun$ (solid line) and $0.5-30~ \msun$
(broken line) in $(\tau_0,\tau_3)$ co-ordinates with $f_a = 40\ \Hz$. Note that the $\tau_0$ should be
multiplied by $(f_a/f_l)^{8/3}$ in order to get the Newtonian chirp time.}
\label{boundary}
\end{figure}

For the mass range of 
$1~ \msun \leq M \leq 30~ \msun$, the longest 
chirp signal occurs when each of the component masses is  equal to $1 ~ \msun$, 
and is  $\sim 95$ sec in duration. 
The data segments are taken to be 512 sec each (greater than three times $\Delta T_{\mathrm{chirp}}$), so $\tpad \sim 417$ sec.
The parameter spaces for the individual mass ranges of $1~ \msun \leq M \leq 30~ \msun$ and 
$0.5~ \msun \leq M \leq 30~ \msun$ in the $\{ \tau_0, \tau_3 \}$ coordinates are shown in 
Fig. \ref{boundary}.

The mass range  is set to $1~ \msun \leq M \leq 30~ \msun$ in this paper and is  typical  for the initial
detectors. The lower mass limit is chosen relatively higher, because higher masses lead 
to higher SNR (the SNR scales as $M^{5/6}$), in order to compensate for the relatively 
lower sensitivity expected for the initial detectors. Choosing the fiducial frequency 
$f_a = 40\ \Hz$, the area of the parameter space is $8.5 \ \sec^2$.  If the parameter 
space is covered by a rectangular lattice of templates, obtained by using the largest inscribed
rectangle within 
the contour of $3\%$ mismatch, the density of templates is $\sim 1300 /\ \sec^2$.
Multiplying the density by the area gives about $\ntemp = 11,050$ templates which are
needed to cover the parameter space. 
One typically takes $512 \ \sec$ data trains sampled at $2048 \ \Hz$ which gives 
$N = 2^{20}$. With a padding length of $ \tpad \sim 417 \ \sec$ and using 
Eq. (\ref{oper}, \ref{cspd}), we see that the online Flop rating required is 
$\sim 3.2$ G-Flops. In addition to  
the cost of FFTs, there are overhead costs, incurred in element by element multiplication of the 
data and template vectors and data preprocessing. However, we neglect them in our analysis since they are small 
compared to the FFT costs. 

The above estimate however does not take into account boundary effects, where the templates 
can spill out of the searched parameter space because of imperfect tiling. This effect can 
considerably increase the number of templates. The current LIGO template placement code \cite{lal} was run for 
the above parameters and was found to generate almost three times as many templates as in this simple 
estimate. Since the current code sets the fiducial frequency equal to the lower cut-off frequency, 
$f_a$ was set at 30 Hz for the purpose of obtaining the number of templates. 

Lowering the mass limit, increases the 
area of the parameter space and thus the computational cost. The results are described 
in Table \ref{tab:1}.

\begin{table}
\caption{Estimates of number of templates and online computational cost for flat search. 
Templates are placed using the method prescribed by Owen and Sathyaprakash \cite{owen2}. 
The prescribed 
minimum mismatch level is set at 0.97. The area of the parameter space is computed for
the fiducial frequency of $f_a = 30 \ \Hz$. 
The upper mass 
limit in all the cases is set to $30\ \msun$.
}
\begin{center}
\begin{tabular}{cccc}
\hline
Minimum Mass Limit & \ \ \ Area\footnotemark\ \ \   & \ \ \ $\ntemp$  & $\nflop$ \\
                   &    $(\sec^2)$     &                 & (G-Flops) \\
\hline \hline
$1.0 \msun$\ \ \ \  & \ \ 29.3   & $\ \ \ 3.34\times 10^4\ \ \ $ & \ \ 10.1\ \  \\
$0.5 \msun$\ \ \ \  & \ 172.9    & $1.64\times 10^5$ & 49.5\\
$0.2 \msun$\ \ \ \  & 1687.4 &   $1.16\times 10^6$ & 350.0 \\ \hline
\label{tab:1}
\footnotetext{The area of the parameter space scales as $f_a^{-13/3}$.}
\end{tabular}
\end{center}
\end{table}

As mentioned before, the template placement code presently  implemented employs effectively a rectangular 
tiling method.  We note here that if a hexagonal closed packing method were used instead, significant 
reduction in the online computational speed can be accomplished, because of 
the efficiency of this method in covering the parameter space is greater than rectangular 
packing \cite{brady,ourpaper,pinto}. As a matter of fact, the hexagonal tiling method gives a 
template density which is about 
$\sim 25\%$ less, thus effectively reducing the required online computational speed by 
the same factor.

\section{Extended hierarchical search}

The flat search described in the previous section is not computationally optimal. This is due to the fact
that a grid of large number of templates is required  in order to densely span $\vec \mu$, satisfying
the criterion that the minimal correlation for two neighboring normalized templates does not fall below a
prescribed value of $0.97$. 

It turns out that we can significantly reduce the computational cost by
constructing two sequential grids of varying template densities instead of one \cite{monty1,monty2}. The philosophy behind 
this approach is the fact that one expects real events to be not only weak but also very rare. This implies 
that most of the time, one is only spending time in sifting through interferometer noise. In this sense, scanning
the parameter space so thoroughly at each observational epoch is wasteful. On the contrary, if
a trigger mechanism exists which would pick up only statistically likely candidates, then a higher parameter
resolution
search (flat search) can be performed {\em{locally}} on pre-screened candidates  possibly containing a
chirp event. Thus figuratively, the hierarchical search method is a layered search. The coarse bank layer is
the trigger which flags a likely event. This is then followed up by the fine grid of templates {\em{a la}} 
flat search method described above. The difference is that only a few fine grid templates can accomplish 
the job of thoroughly verifying the claim of the trigger stage coarse grid templates. The computational 
cost thus is reduced as  fewer templates are employed in the overall scheme. 

The above idea can be extended even further to include the time of arrival of the signal in the overall
scheme of hierarchy as we shall now explain. From (\ref{eq:htilde}), one notices that the signal power
spectrum  is a decreasing power law in frequency :
\be
\left | \htilde \right | ^2 \propto f^{-7/3}.
\label{eq:powspec}
\ee
From (\ref{eq:powspec}), we see that most of the signal power is contributed from the  lower frequency
bands. An immediate consequence is that one can cut off the chirp at a lower cut-off frequency and re-sample
it at a lower Nyquist rate without losing too much power in the signal. In the EHS we propose to sample the
trigger stage data train at a much lower sampling rate than that used in a flat search algorithm. As a
proof of principle, we notice that the fractional signal to noise ratio $\rho$ as a function of 
an arbitrary upper cut-off frequency $\fcut$ is given by $\rho(\fcut) = I(\fcut)/I(f_c)$, where
\be
I(\fcut) = 2 \int_{f_l}^{\fcut} \frac{df}{f^{7/3} S_n(f)}, \ \ \ f_l < \fcut \leq f_c.
\label{eq:rhobar}
\ee
If the ISCO cut-off is used, $\fcut$ must be replaced by the appropriate plunge cut-off 
frequency. However, this is applicable for high masses in the first stage 
$(M \geq 16 \ \msun)$, which is a tiny fraction of the parameter space - implying a 
negligible effect on the overall gain factor. 
A significant fraction of the SNR
can be recovered from a relatively small value of $\fcut$. Specifically, for $\fcut = 256 \ \Hz$, 
(i.e. a sampling rate of $\fsamp^{(1)} = 512 \ \Hz$) almost $92\%$ of the signal power is 
recovered. The most obvious advantage of lowering the Nyquist rate in the 
first stage is that we have to contend with fewer points in computing the
FFTs, (the cost of which scales as $N\log_2 N$), leading to a reduction in first stage 
computational cost. For example, a chirp cut-off at $\fcut = 256 \ \Hz$ reduces the cost 
of FFT by almost a factor of $4$. 

The equation (\ref{eq:rhobar}) correctly estimates the loss in SNR if the time of arrival 
parameter $t_a$ of the signal coincides exactly with a sampled time in the time-series. If, 
as will be the case in general, $t_a$ is not a sampled time, then there will be an additional 
loss of SNR due to mismatch in time of arrival parameter. Considering the `worst case' scenario 
when $t_a$ of the signal lies exactly between two consecutive time samples, we numerically 
estimated this additional loss in SNR in two different cases : (a) when the chirp times of the 
signal match exactly with those of a template and (b) when they lie exactly in between two 
neighboring templates. In the former case, we found that the additional loss in SNR is less 
than $5\%$ and in the latter case, a little more than $2\%$. It is to be noted however, that 
our template placement schema (see Sec \ref{sec:templateplacement}) introduces some overlap 
between neighboring templates, which effectively reduces the maximum mismatch. We find that  
this adequately compensates for the additional loss in SNR and no further reduction of first 
stage threshold (prescribed in Sec \ref{sec:chooseZetaOne}) is necessary.
 
In the earlier method given by Mohanty and Dhurandhar \cite{monty1}, the hierarchy was on the two mass parameters alone
in the sense that a coarse template bank spanning the mass parameters was used as the trigger, followed up
by a fine template bank. Analogously in the EHS algorithm, we not only retain this hierarchy of templates spanning the mass
parameters but also extend it to the time of arrival by coarser sampling of the signal  at the
trigger stage. It is in this sense that  the EHS should be thought of as a three dimensional hierarchical
search.

But quite aside from this, lowering the Nyquist rate also affects the contours of the ambiguity function 
$\amb$ which in turn affect the template coverage in the first stage. Also a lowered
signal power in the first stage will require lowering the trigger threshold. This will affect the false alarm rate due to
noise in the interferometer and in turn the overall computational cost. In the next few subsections,
we systematically address these issues and prescribe the optimum method of implementing the EHS algorithm.

\subsection{Setting up the fine bank}
\label{sec:finebank}

In this subsection, we will discuss the setting up of the fine bank templates and the second stage threshold.

The templates in the second stage or the fine bank stage of the EHS is set up in a way that is identical to 
the flat search method discussed in the earlier section.  The ambiguity function 
$\amb$ centered at the $\vec \mu$ is defined as the correlation between two neighboring normalized templates whose mass parameters 
differ by $\Delta \vec \mu$:
\be
\amb (\vec \mu, \Delta \vec \mu) = \max_{t_0} \left ( \tilde{h} (\vec \mu,\vec \lambda), \tilde{h} (\vec \mu + \Delta
\vec \mu,\vec \lambda) \right ).
\label{eq:amb}
\ee
The contours of $\amb$ are closed curves and the area enclosed within this curve will be frequently referred 
to as a {\em{tile}}. The shape of this tile depends on the  contour level $\Gamma$ where $0 < \Gamma < 1$. 
For values of $\Gamma$ close to 1, the tiles are elliptical in shape. However, as $\Gamma$ 
significantly differs from unity, the shape can be 
quite irregular. This is important from the point of view of template placement where the parameter space 
must be efficiently covered by these tiles. In the fine bank, the templates are laid 
out in such a way that nowhere does $\Gamma$ fall below $0.97$. For such high values of 
$\Gamma$, the ambiguity function can be approximated quite
well by a Taylor expansion of $\amb$ truncated after the quadratic term. This allows analytic 
calculation of the dimensions of these elliptical tiles and placement of templates 
\cite{owen1,owen2} over $\vec \mu$. This is not true,  however, for smaller values of the match 
level.  The number 
of templates indirectly
affect the minimum observable signal strength in the second stage of the hierarchical search method
and thus the overall minimum detectable signal strength.

In order to proceed with the numerical analysis, the noise in the interferometers is assumed to be a stationary, 
zero-mean Gaussian random process. 
Thus the integrated signal to noise ratio $\rho$ calculated using 
(\ref{eq:rho}) should also be considered to be a realization of a random process whose 
probability distribution depends on the presence or absence of the signal. In the presence of a signal, 
the distribution is a Rician, whereas in
the absence of the signal, $\rho$ is Raleigh distributed. The detection strategy 
is to set a threshold $\zeta$ such that given one instance of $\rho$, a decision can be made 
with some statistical significance, whether or not, it was drawn from a Raleigh or a Rice 
distribution - which would correspondingly imply the presence or absence  of a GW signal. 

Given a sufficiently high $\zeta_2$ (in the second stage), even in the absence of a signal, there may be some
crossings of $\rho$ above this value leading us to incorrectly deduce the presence of a signal. This threshold is thus
primarily set such that the probability of these false alarms is no more than once per year. 
Thus, $\zeta_2$ is set at
\be
\zeta_2 = \sqrt{ 2 \ln (\epsilon \ntemp^{(2)} T_{\mathrm{yr}} \fsamp)},
\label{eq:zeta2}
\ee
where, the data train is sampled at $\fsamp$ Hz and $T_{\mathrm{yr}} = 3.15\times 10^7$ 
are the seconds in a year. The factor
$\epsilon$ arises from the correlation between templates and correlations 
between successive time-lags in the filtered output \cite{schutz}. Because the chirp filter 
is a narrow band filter and band-passes the data, the inverse Fourier transformed data in the time
domain becomes correlated as it is essentially generated from few frequency bins. Thus 
$\epsilon$ gives the fraction of the number of independent Gaussian random variables in the 
total filter bank output. From our simulations we
find that $\epsilon \sim 0.6$ in the second stage where the templates are closely spaced, 
and $\epsilon \sim 0.9$ in
the first stage where the templates are farther apart. If there had been no correlation 
between the templates, we could have set
$\epsilon=1$. Using Eq.(\ref{eq:zeta2}), and the fact that the sampling frequency in the second  
stage is $2048\
\Hz$, we find that the second stage threshold must be put at $\zeta_2 = 8.2$. If this threshold coincides with
the strength of minimum observable signal $\sobs$, then the detection probability is just $50\%$. 
In order to achieve higher detection probability, the minimum observed 
strength of the signal should be somewhat greater than $\zeta_2$. Following 
\cite{monty2}, we compute the detection probability by considering the joint probability of 
two neighboring templates around the signal. Then for attaining a minimum of $95\%$ 
detection probability, the minimum strength of the signal should be 
$\sobs = \zeta_2 + \Delta \zeta$ where $\Delta \zeta \sim 0.7$.
The second stage templates can thus detect a
signal of strength $\sobs = 8.9$ with a detection probability of $95\%$ or greater. From the 
minimum match that we have chosen $(97\%)$, we see that this corresponds 
to a signal of minimum strength $\smin = 9.2$ arriving at the detector.

\subsection{Setting up the trigger stage templates}

The data trains in the trigger stage of the extended hierarchical search are sampled at $512\ \Hz$, and the template chirps are 
terminated at $256\ \Hz$, so as to avoid power aliasing effects in the FFTs. These templates are used as triggers. 
The neighborhood of a 
trigger is then searched using the second stage fine bank templates. 

We begin by describing the contours of $\amb$ when the GW chirp is  cut-off at a lower 
frequency, i.e. when $\fcut = 256 \ \Hz$. The contours at different $\Gamma$ levels are not 
only irregular in shape, but cover larger tile areas compared to those with cut-off  
$\fcut = 800\ \Hz$.  The basic reason for the 
irregular shape of tiles is that, the higher order post-Newtonian (PN) terms that 
appear in the phase function $\Psi$ in Eq. (\ref{eq:phase}) decay slowly with the frequency $f$. 
The phase function $\Psi$ appears implicitly in the integral of the ambiguity function.
While the Newtonian term (the $\tau_0$ term) falls off as $f^{-5/3}$, the $\tau_2, \tau_3, 
 \tau_4$ (1 PN , 1.5 PN, 2 PN) terms fall off as $f^{-1}, f^{-2/3}, f^{-1/3}$ respectively. 
The 1.5 PN and 2 PN terms contribute to the phase significantly at high 
frequencies since their fall off is relatively slower. Thus an upper cut-off of 256 Hz 
verses 1024 Hz produces tiles that  differ significantly in shape. 
The irregular and asymmetric shape of the contour can also be viewed as an effect of the
curvature of the manifold. Because the 2 PN or the $\tau_4$ term has not decayed, it contributes 
to the metric  making it non-constant in the $\{ \tau_0, \tau_3 \}$ space.
The bigger tiles are essentially due to the fact that, it is the higher frequencies which 
resolve between the masses, and if these are cut off,  the ambiguity function decays 
more gradually, resulting in bigger tiles. At $\Gamma \sim 0.8$ the tiles with $\fcut = 256$ Hz 
are larger in area than those with $\fcut = 800$ Hz by about a factor of 2.  Moreover, 
because the contours are large, non-local effects are important, compounding the curvature 
effects and adding to the irregular shape of the contour.

In the earlier hierarchical scheme, the ambiguity function in both the trigger and fine bank stages were the
same. The trigger stage templates were constructed by letting the trigger threshold drop from $\zeta_2$ to a 
lower value $\zeta_1$ such that a balance was struck between the following opposing forces :
$\zeta_1$ was low enough to allow larger tiles to cover the parameter space with fewer templates and thus reduce the
computational cost, but high enough so that false alarm crossings due to noise do not compromise the cost
advantage by increasing the second stage cost. Signal power in the trigger stage was the same as in the fine
bank stage since the sampling rate was the same in both stages.

In the EHS implementation of the hierarchical search, we are faced with a different proposition - we have
reduced signal power in the first stage due to lower cut-off frequency. Also, we have a 
different $\amb$ in the two stages. Given these differences, we need to devise how to set up the coarse 
grid of templates. We now describe the template placement for the first stage of the  
EHS algorithm. 

\subsection{Setting up the trigger stage tiles}
\label{sec:templateplacement}

The tiling problem stated from an operational point of view is the following :
given the  prescribed match level $\Gamma$ for the trigger stage,
we need to tile the parameter space {\it efficiently} with closed contours of $\amb$. 
The following conditions must be satisfied \cite{pinto} for efficient template 
placement: 
firstly, minimal mismatch must be satisfied {\it i.e.} 
${\mathcal{H}} \geq \Gamma$ inside and on the contour. Also, templates should be placed such that there 
are no uncovered spaces (no holes) and there is minimal overlap between templates. 
For computational ease, the templates should preferably lie on a regular grid or 
at least a piece-wise regular grid. As stated earlier, the ambiguity function ${\mathcal{H}}$ is quite wide at 
low frequency cut-offs since the mass resolution is poor in absence of high 
frequency components. We choose $\Gamma = 0.8$ contours in the trigger stage to tile the available 
parameter space, since the total cost is minimum for this value of $\Gamma$. The 
results quoted at the end of the paper are also for other values of $\Gamma$ but close to 0.8.  
It is quite unwieldy 
to use these contours {\it prima facie} to tile the parameter space. A possible 
solution is to carve a regular polygon - e.g. a rectangle out of these contours and 
use them as tiling blocks. To minimize the computational cost, we must choose the 
rectangle as large as possible. The details of a template placement (tiling) 
algorithm are now described. The demands made upon the tiling algorithm are several - 
(i) given a $\Gamma = 0.8$ contour centered at some point in the parameter 
      space, 
      the largest rectangle should be identified {\it automatically} and used as the 
      representative tile at that point,
(ii)  having placed a template at a point in the $(\tau_0,\tau_3)$ space, 
      the optimum position of the neighboring templates should be estimated such 
      that the template placement conditions are  satisfied, and lastly
(iii) the parameter space in $(\tau_0,\tau_3)$ co-ordinates has non-trivial 
curvature because of which the contours are asymmetric and also have differential rotation as a function  of
$(\tau_0,\tau_3)$ and one 
cannot set up a global regular grid 
for  arbitrary values of $\Gamma$. The tiling scheme should  correct for these 
differential rotations  so that there are no uncovered spaces due 
to differential rotation of the rectangular tiles.
The description of this tiling algorithm proceeds in the following logical sequence - 
we first describe a method of 
carving a large rectangle given a contour of ${\mathcal{H}}$ at a specified value of 
$\Gamma$. Thereafter a method for placing neighboring templates is described. This method 
automatically takes into account the differential rotation of the templates. Finally, 
an overall placement schema is given which involves `stacking'.
  
\begin{figure}[h]
\centering
\includegraphics[width=0.75\textwidth]{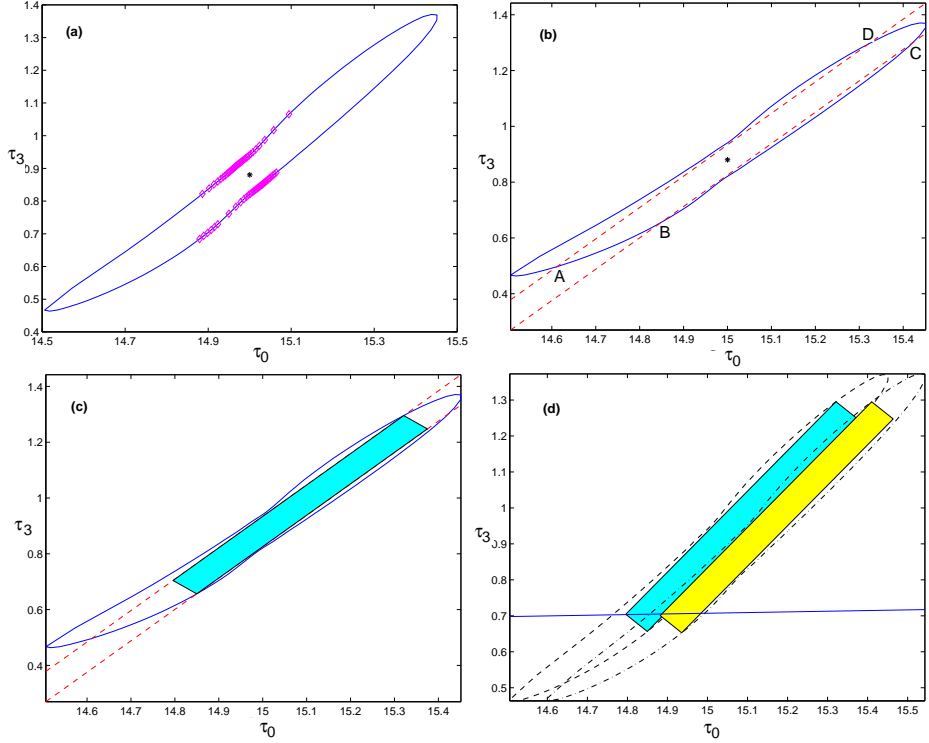}
\caption{This figure illustrates the sequential steps for obtaining the inscribed rectangle and placing the
templates. Units of $\tau_0$ and $\tau_3$ are in seconds. (a) The orientation of the rectangle is first determined by fitting straight 
lines to points around the concave sides of the contour. (b) Then lines 
with slope $m = \tan \theta$  are drawn inside the contour on either side. These 
intersect at four points A,B,C,D. (c) The  rectangle is then extracted from the four 
points by dropping perpendiculars. (d) Figure illustrating the placement of two neighboring 
templates along some curve $f(\tau_0,\tau_3) = 0$ parallel to the $\eta = 1/4$ line. 
Note the slight overlap between the templates due to $\dth=0.025$ correction and 
the negligible spillage of rectangle below the equal mass line. }
\label{fig:combined}
\end{figure}

The steps involved in  identifying a sufficiently large rectangle given a $\Gamma = 0.8$ contour 
is illustrated using one centered at $(\tau_0 = 15.0 \ {\mathrm{sec}}, \tau_3 = 0.88 \ {\mathrm{sec}})$.
The rectangle must align itself along the length of the contour. So the first 
step is to identify the orientation of the rectangle. This is done by fitting 
straight lines to the concave sides of the contour with respect to the center. The 
points are so chosen that it also includes the one which is radially closest to the 
center as can be seen in Fig \ref{fig:combined}(a). If the slopes of these two lines are $
\tan \theta_1$ and $\tan \theta_2$ respectively, then the side of the rectangle is 
chosen to have the average slope $\tan \theta$: 
\begin{equation}
\tan \theta = \frac{1}{2} (\tan \theta_1 + \tan \theta_2). 
\end{equation}
The next step involves drawing two parallel lines with their slope 
$\tan \theta$, such that they (a) are as far out as possible from the center 
(b) intersect the contour at only two points (c) lie completely inside the 
contour. This is clearly shown in  Fig \ref{fig:combined}(b) where the 
intersection points are labeled as A,B,C and D.

Two adjustable parameters are used to fix the distance of each side of the rectangle from the center of the contour. These parameters should be chosen such that 
conditions given above  are satisfied. Due to the peculiar shape 
of the contour, the lines (see Fig \ref{fig:combined}(b)) may intersect the contour at 
multiple points. This must be avoided by fine tuning the parameters.

The final step  is to read off the co-ordinates of the rectangle 
by dropping suitable perpendiculars. This is shown in  
Fig \ref{fig:combined}(c), and the resultant rectangle is shaded. In general,  if the axes are scaled
differently, the rectangles would look like  parallelograms. 

By following the above steps,  a sufficiently large rectangle can be identified from within 
any prescribed contour. Before placing neighboring templates, one needs to correct for the 
differential rotation of the templates. This 
is achieved with an  adjustable angular parameter $\dth$ in the schema which 
deliberately induces some overlap between nearest neighbor templates by reducing the 
{\it effective} length and breadth of the rectangle that is used. 
If the actual dimensions of the rectangle are $l$ and $b$, the $\dth$-corrected values 
$l'$ and $b'$ are given by
\begin{eqnarray}
b' &=& \ell \,\tan \left [ \tan^{-1}(\frac{b}{\ell} ) - \dth \right ], \\
l' &=&    b \,\tan \left [ \tan^{-1}(\frac{\ell}{b} ) - \dth \right ].
\end{eqnarray}
The parameter $\dth$ should be chosen very carefully so that minimal overlap condition is 
satisfied.
\begin{figure}[h]
\centering
\includegraphics[width=0.45\textwidth]{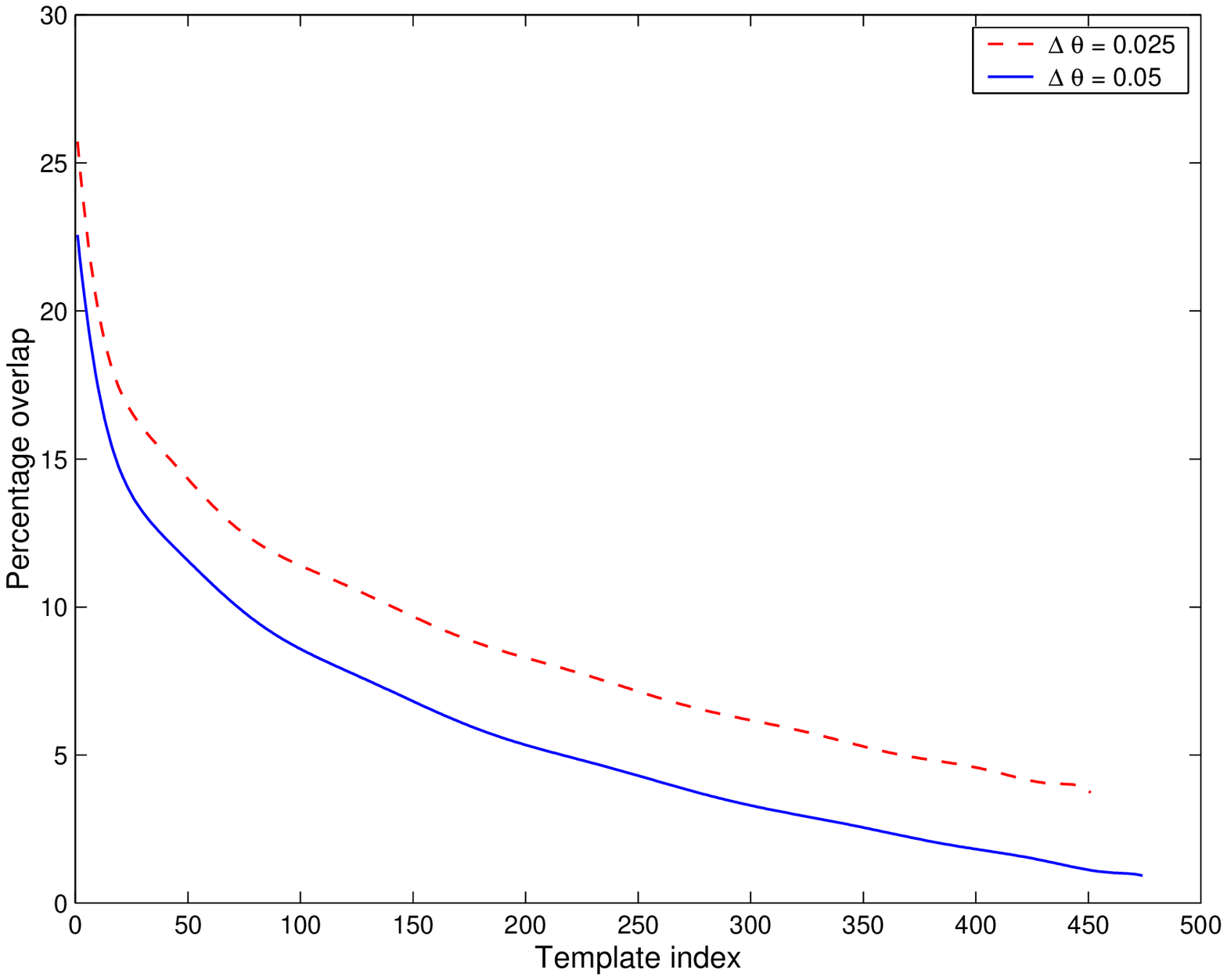}
\includegraphics[width=0.45\textwidth]{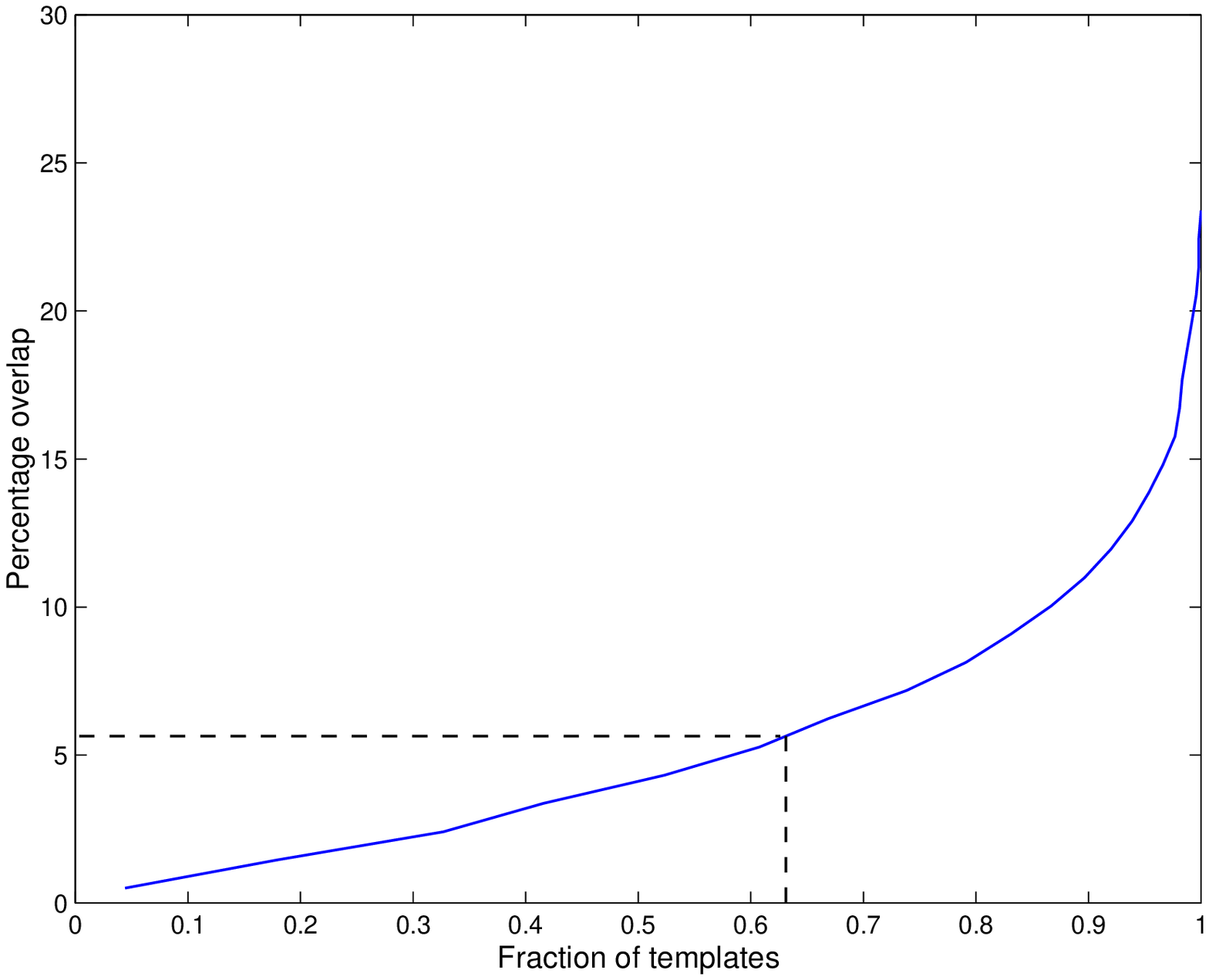}
\caption{(a) Percentage overlap between neighboring templates after 
$\dth$-correction. The values are $\dth = 0.025$ (top) and $\dth = 0.05$ 
(bottom). Percentage overlap is defined as the ratio of area common to two neighboring templates to the area of any one of them
expressed as a percentage. Note that fewer templates are required for smaller $\dth$ parameters. (b) Cumulative distribution of templates according to the $\dth$-induced  
area overlaps between neighbors for $\dth = 0.025$. The average percentage overlap in this case is $5.7\%$. 
The dotted line shows that $\sim 63\%$ of the templates have percentage overlaps less than this average value. 
}
\label{fig3}
\end{figure} 
 The curves
in Fig \ref{fig3}(a) show the overlap fraction between neighboring rectangles 
as we move along the $\tau_0$ axis for $\dth = 0.05$ and $\dth = 0.025$ 
respectively.
We need to tweak the parameters such that the overlap 
approaches zero as quickly as possible without ever going below it. This will 
guarantee  a minimum number of templates. Fig. \ref{fig3}(b) shows that for 
$\dth = 0.025$, most of the templates have neighboring overlaps of $\sim 5\%$ of 
their width.

One would like to place the templates along some curve $f(\tau_0,\tau_3) = 0$. For
example, this curve could be the equal mass line $(\eta = 1/4)$. Once the
curve is chosen, the center of the neighboring template is found by moving out by a 
distance
$\Delta \tau_0$ along the $\tau_0$ axis given by 
\begin{equation}
\Delta \tau_0 = b'\,\frac{\cos \beta}{\sin (\theta-\beta)},
\end{equation}
where, $b'$ is the $\dth$-corrected breadth and the local slope of the curve 
$f(\tau_0,\tau_3) = 0$ is $\tan \beta$. For the $\eta = 1/4$ line ($m_1 = m_2$), the 
slope can be analytically determined :
\begin{eqnarray}
\tan \beta &=& \frac{d \tau_3/d M}{d \tau_0/dM},  \nonumber \\
           &=& \frac{64}{25} \pi^2 f_a M,
\end{eqnarray}
where $2~ \msun\leq M \leq 60~ \msun$ is the total mass of the compact binary system. 
The $\tau_0$ co-ordinate of the center of the new rectangle is given by 
$\tau_0 + \Delta \tau_0$. The $\tau_3$ co-ordinate can be found from the 
equation of the curve. 

This is a {\it dynamic} method of placing the templates. 
However, we do not go to such lengths in placing the neighbors. 
Since the average orientation of the rectangle  is known, a constant fudge-factor in 
multiplying $b'$ is quite sufficient. In this case, one relies entirely on the 
dynamical change in $b'$ across the parameter space. Choosing  
$\Delta \tau_0 \simeq 1.28b'$ is sufficient for our purposes. This choice will slightly 
increase the first stage cost.
The placement of two neighboring 
templates is shown in Figure \ref{fig:combined}(d). Note that the overlap between the rectangles 
is small.

\begin{figure}[h]
\centering
\includegraphics[width=0.45\textwidth]{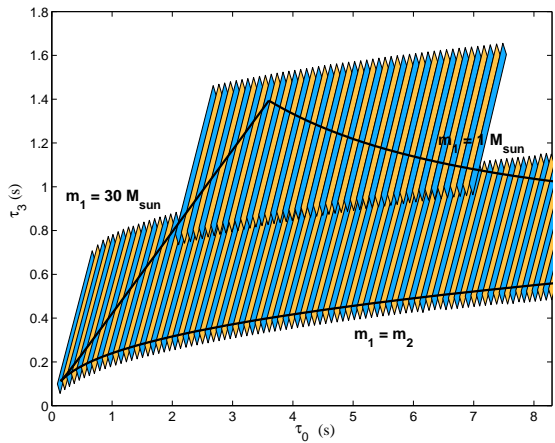}
\caption{Figure illustrating the 
stacks of templates on top of each other covering a
part of the targeted parameter space. The dark lines are the boundaries of the parameter space for the 
mass range $1~\msun \leq m_1, m_2 \leq 30~ \msun$. Due to the large size of the rectangles, the hierarchical search will
actually cover more of the parameter space than the flat search it replaces. Note that this figure is a zoomed view of the first
$20\%$ of the $\tau_0$ domain.}
\label{fig:stacks}
\end{figure}

Our overall  strategy of template placement is to build {\it stacks} of 
templates as can be seen in 
Fig \ref{fig:stacks}. Here it is worth noting that all earlier schemes \cite{grasp,pinto} 
start by placing templates on the equal mass 
($\eta = 1/4$) line. However, the space below this line is unphysical - 
as such, almost $50 \%$ of the templates centered  on this line will be 
wasted and  the final result is that we end up using too many of first stage templates 
in covering the parameter space. We can ensure more efficient coverage
by constructing templates on a line parallel
to the $\eta = 1/4$ line but which is offset by parameters $\ell_p$
and $\theta$ as : 
\begin{eqnarray}
 \tau_0 &=& \tau_{0,\eta=1/4} + \ell_p \cos \theta, \\
 \tau_3 &=& \tau_{3,\eta=1/4} + \ell_p \sin \theta,
\end{eqnarray}
where $\ell_p$ is so chosen that the rectangle constructed spills minimally below 
the equal mass line.  Due to  the asymmetry of the contour, 
$\ell_p$ need not be equal to half the length of the inscribed rectangle.

By shifting up the templates in this way, a larger area of the 
parameter space is covered than if their centers were placed on the equal 
mass line. Even though one needs to place templates all along the curve for 
the first stack - for the subsequent stack, templates need to be placed only 
between points of intersection of the top of the templates of the earlier 
stack with the parameter space boundaries 
$m_1 = M_{\mathrm{max}}$ and $m_1 = M_{\mathrm{min}}$. 
Thus the first stack completely covers the low mass end of the parameter space 
from $\tau_0 \gsim 7$ sec. Also the high mass end of the parameter 
space, $\tau_0 \lsim 2 $sec is covered by the first stack.   
Having placed the first stack of templates in this way, the next stack is 
constructed by choosing the curve that is parallel to the first stack, but 
offset further according to the sizes of the templates in the previous 
stacks and adjusted for minimal overlap. The two stacks completely cover the parameter 
space. We find that the total number of coarse templates required is 
$\ntemp^{(1)} \sim 510$ for the contour level of $\Gamma = 0.8$.
 
If the lower mass limit had been smaller, say, $0.5~ \msun$, then more than two stacks 
would have been required. It is easy to see how this schema can be generalized for 
smaller component masses. 
The details of placement using the above scheme towards the high 
mass end of the parameter space is shown in Figure \ref{fig:stacks}.

The template placement schema described above {\em stacks} rectangular templates centered on  curves parallel to the equal mass line, 
and  the size of each tile covers a significant
portion of the parameter space $\vec \mu$. As such,
it is important that the length of templates in any stack does not vary too much. Indeed, the length of these tiles is seen 
to change by only a few percent over $\vec \mu$. However their orientations change by $\sim  0.1$ radians and is seen to be a 
dominant effect at the high mass end. The parameter $\dth$ takes care of this in our schema by introducing extra overlaps between 
neighboring templates. We thus find that the issue of varying template sizes is 
less of a concern than varying orientation of the templates. 

This completes the formal description of the template placement algorithm.

\subsection{Simulation of EHS}
\label{sec:simu}

We performed Monte Carlo simulations to check the performance of the EHS algorithm against 
simulated detector noise with injected signals at different positions in the
parameter space. 

Only the trigger stage of the EHS is of relevance in this
context since, the flat search used locally around a candidate event in the 
second stage is quite efficient in extracting relevant signals. The purpose of these 
simulations was to ascertain if the noise crossings were as expected as used in the analysis 
and design phase of the EHS algorithm. Secondly, since template placement method used in the 
first stage does not center the tiles on the $\eta = 1/4$ line, we
would like to check how efficiently signals with parameters close to this line are triggered 
by the first stage templates. 

We chose a bank of 
$\ntemp^{(1)} = 11$ trigger stage templates covering the fraction of the parameter space
where the ambiguity function is non zero. For 
ease of discussion, we label these from $T_1$ to $T_{11}$ (see Figure
\ref{fig:sig+temp}).
The simulation was carried out in two phases. In the first phase we parsed 
data segments of $512 \ \sec$ duration containing only simulated Gaussian stationary noise 
with LIGO-I power spectral density through these templates. In the second phase, a chirp 
signal with a predetermined maximum SNR $\rho$ was added to the 
noise. The trigger stage sampling rate was chosen to  be $512 \ \Hz$, such that the number 
of points in each data segment in this stage was $N_1 = 2^{18}$. The spectral window for this stage is 
taken to be $30\ \Hz \leq f \leq 256 \ \Hz$. 

\begin{figure}[hbt!]
\centering
\includegraphics[width=0.49\textwidth]{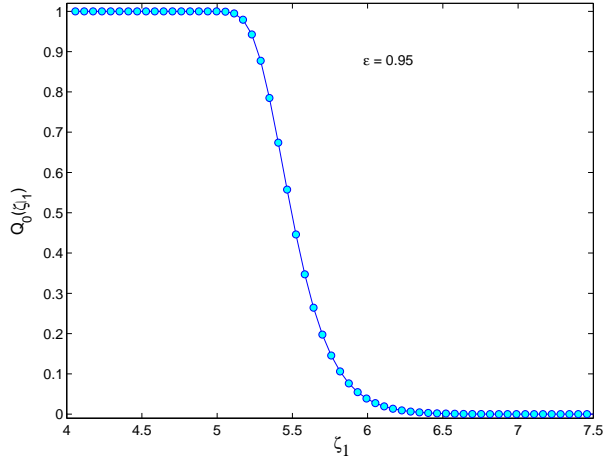}
\caption{False alarm probability  estimates as a function of the first stage threshold for $N_T = 11$ templates. The solid 
curve is obtained from (\ref{eq:Q0}) with $\epsilon = 0.95$, whereas the solid circles are 
the Monte Carlo estimates. }
\label{fig:Q0}
\end{figure} 

The noise in the interferometers is assumed to be a Gaussian, whose characteristics remain constant in 
time. We used the LIGO-I noise spectral density to color samples of white Gaussian noise. 
From (\ref{eq:rho}), one notices that in presence of noise only, the filtered SNR from each 
of $\ntemp^{(1)} = 11$ templates is  expected to be 
realizations of a random process with Raleigh distribution. 
As remarked earlier in the context of fine bank placement, the filtered output from each of the templates is correlated. 
We account for this fact by introducing the quantity $0 < \epsilon < 1$, which is used to reduce the 
effective number of independent random variables. Using these, one can calculate the probability of the test statistic  
crossing a preset threshold $\zeta_1$. This is precisely the false alarm probability 
$Q_0(\ntemp^{(1)},\zeta_1)$ as a function of the number of templates used and the first stage 
threshold. Assuming Gaussian noise, it can be expressed as
\be
Q_0 (\ntemp^{(1)},\zeta_1) = 1 - \exp[\epsilon \ntemp^{(1)} \npad \exp( - \frac{\zeta_1^2}{2} )],
\label{eq:Q0}
\ee
where $\npad = \tpad \times \fsamp^{(1)}$ is the number of points in the non-overlapping segments of adjacent 
analysis epoch,
$\tpad$. This curve is then compared with the result obtained from simulations and 
 $\epsilon$ is chosen to make the analytical expression agree with the Monte Carlo results. 
The parameter $\epsilon$ should be taken as a measure of the statistical 
independence of the templates; $\epsilon = 1$ means all the $\ntemp^{(1)} \npad$ random variables 
are independent. In Figure \ref{fig:Q0}, we have plotted the Monte-Carlo estimates of the 
false alarm probability. For the trigger stage of the EHS, where the
distances between templates are quite large, we expect the assumption of statistical 
independence to hold to a large extent. Indeed, the best fit parameters  for the curve 
gives $\epsilon \simeq 0.95$.

\begin{figure}[hbt!]
\centering
\includegraphics[width=0.49\textwidth]{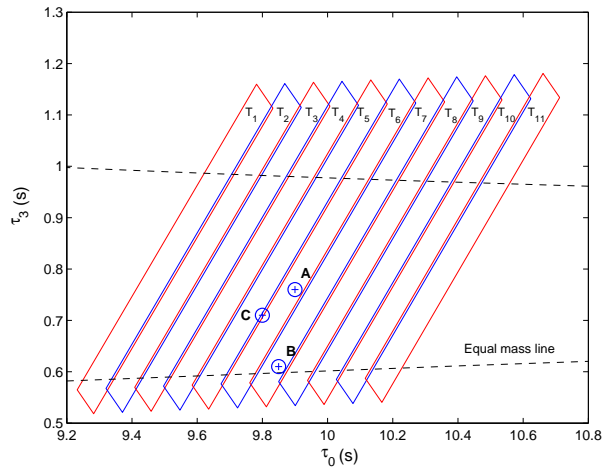}
\caption{The positions of $\ntemp^{(1)} = 11$ templates used in the simulation and the relative positions of the signals
injected into the interferometer noise. The templates are labeled $T_1$ to $T_{11}$. See the text for a description of the three
simulation cases A, B and C.}
\label{fig:sig+temp}
\end{figure} 

In the next phase of simulations, we added known signals to interferometer noise realizations and 
processed the data through  eleven templates.

We chose our signal parameters at three 
characteristic positions in the parameter space with respect to the centers of the templates 
such that they represented three different detection scenarios. In  case (A), the signal 
parameters were chosen to be favorably placed with respect to the filters. This was done by 
taking a point close to the center of $T_6$ (see discussion in caption of Fig
\ref{fig:sig+temp}). In case (B), we chose the signal to have 
parameters such that it lay in the area bounded by $T_7$, but placed far away from the 
template centers - close to the equal mass line. In case (C), the signal was chosen to lie 
close to the equal mass line, exactly between $T_5$ and $T_6$. We believe that these scenarios are 
typical for templates placed anywhere in the $(\tau_0,\tau_3)$ parameter space, as it is more or less
flat.

\begin{table}
\caption{Monte Carlo estimates of detection probability and percentage of triggers generated by 
individual $N_T = 11$ templates in three
different cases : (A) signal lies on $T_6$ and close to template center (B)
signal parameters lie on $T_7$ and are quite far away from the center of the templates. Actually in this case the parameters
lie close to the equal mass line. (C) The parameters lie exactly between $T_5$ and $T_6$, and close to
the equal mass line. The templates are individually labeled from $T_1$ to $T_{11}$.}
\begin{center}
\begin{tabular}{cr|rrrrrrrrrrr}
\hline
Case & $Q_d (\%)$  & $T_1$ & $T_2$ & $T_3$ & $T_4$ & $T_5$ & $T_6$ & $T_7$ & $T_8$ & $T_9$ & $T_{10}$ & $T_{11}$ \\   
\hline \hline
A & $99.1 $ & 0.0 & 0.0 & 0.0 & 0.0 & 0.3 & 99.2 & 0.5 & 0.0 & 0.0 & 0.0 & 0.0 \\
B & $97.2 $ & \ 0.0 & \ 0.0 & \ 0.0 & \ 0.0 & \ 0.0 & \ 0.0 & 90.0 & \ 9.9 & \ 0.1 & \ 0.0 & \ 0.0\\
C & $95.7 $ & \ 0.0 & \ 0.0 & \ 0.0 & \ 0.2 & 53.9 & 44.3 & \ 1.2 & \ 0.2 & \ 0.2 & \ 0.0 & \ 0.0\\
\hline
Noise & \ 3.7 & \ 0.3 & \ 0.7 & \ 0.3 & \ 0.3 & \ 0.3 & \ 0.2 & \ 0.2 & \ 0.5 & \ 0.2 & \
0.2 & \ 0.5\\  
\hline
\end{tabular}
\label{tab:2}
\end{center}
\end{table}

When only noise was processed, the trigger stage extended hierarchical search templates registered on an 
average $0.3\%$ 
false alarm probability per template. However we found that these alarms were isolated 
events in the sense that they were not triggered by multiple 
adjacent templates in unison. On the contrary when a bona fide signal was present, there were always a 
cluster of nearby of templates which showed crossings above the trigger stage threshold. For the purposes  
of this paper, we choose the crossing with the maximum SNR for determining the trigger 
template; that is we select just {\em one} template in a given realization of the interferometer noise. 
This is like a one point estimate of the signal parameters when multiple trigger 
templates have shown crossings. The Monte Carlo simulation described below justifies this 
approach. 

The results of this phase of the Monte Carlo simulation are summarized in Table \ref{tab:2}. 
In absence of a signal, we find that all the templates are equally likely to register a 
false alarm, the probability of which is in agreement with (\ref{eq:Q0}). In presence of a 
genuine signal, one notices that when the overlap between the signal and templates is high 
(case A), the signal is almost always correctly triggered (by $T_6$). When the distance 
between the signal and templates is large (case B) then there is significant loss in 
correlation. In this case, we notice that about $10\%$ of the triggers {\em spill over} to 
the adjacent templates. Finally, when the signal lies 
exactly between two templates (case C), the triggers are equally distributed between the 
adjacent templates. 

Given these results, we can draw a strategy for chalking out the area over which the local 
flat search needs to be carried out in the second stage around every first stage trigger 
template. We introduce a factor $\alpha$ which quantifies this idea. $\alpha$ is the ratio 
of the area of the parameter space searched in the second stage for a given click to the area of the 
first stage template. Thus, $\alpha =1$ means the fine bank search is carried 
out within an area of the first stage template. But as the simulations show we need to 
search beyond this region with the fine bank, and thus $\alpha$ needs to be chosen greater 
than unity in most of the parameter space. The parameter space boundaries can however make 
$\alpha < 1$ especially at the low mass end where the parameter space tapers and very 
little area within the parameter space is required to be searched. Below we quote the 
the value of $\alpha$ averaged over the entire parameter space. 
If the interferometer noise is adequately described by a Gaussian random process, then our simulations 
bear out that searching two adjacent templates around the clicked template(s) should account 
for the correct location of the signal to $< 1\%$ error. Thus it is sufficient to set 
$\alpha = 2.5$. 

In presence of non-Gaussian features in the interferometer noise, the behavior of triggers can be quite
unpredictable and we anticipate formidable difficulties in choosing the neighborhood for the 
fine search. Strategy based on the actual noise behavior will have to be devised. More work 
needs to be done towards a robust estimation of signal parameters from the trigger stage 
output itself so that the cost of the second stage can be minimized.

\section{Computational cost of EHS algorithm}

The computational cost in the EHS algorithm comes from (a) processing the data segment through all the
first stage trigger templates, (b) a fine bank search in the neighborhood of all the 
triggers generated in (a) above and (c) cost of preprocessing and passing putative events between coarse and fine
banks of the hierarchical search. In the cost analysis, we do not consider (c). However, we pause to note that if 
(c) is sufficiently large, it can  completely nullify the gains we claim the extended hierarchical search method
makes over a flat search. It maybe possible to setup the problem such that in the worst case, it is no worse than a
flat search.

Let a data segment of $T \ \sec$ duration be sampled at $\fsamp^{(1)} \ \Hz$ in the first stage and 
$\fsamp^{(2)} \ \Hz$ in the second stage. The total floating point operations performed in 
the first stage is given by
\be
\nop^{(1)} \simeq  \ntemp^{(1)} ~(6 \fsamp^{(1)} T \log_2 (\fsamp^{(1)} T),
\ee
where, $\ntemp^{(1)}$ is the total number of first stage templates employed. Furthermore, 
the frequency of  triggers generated in (a) depends on the first stage trigger threshold 
$\zeta_1$. As we shall see later, $\zeta_1 \simeq 6.05$ is the optimum first stage threshold. 
This corresponds to the optimum size of first stage tiles at $\Gamma \simeq 0.82$ contour 
level, which gives us $\ntemp^{(1)} = 560$ in this 
stage. If such a threshold is put, the average number of crossings $\bar{n}_c$  due to noise 
alone (assumed Gaussian and stationary ) is given by
\be
\bar{n}_c \sim \ntemp^{(1)} ~ Q_0(1,\zeta_1),
\label{eq:nc}
\ee
where $Q_0(1,\zeta_1)$ is the false alarm probability for a single template and is given by,
\be
Q_0 (1,\zeta_1) = 1 - \exp[\epsilon N_p \exp( - \frac{\zeta_1^2}{2} )],
\label{eq:Q01}
\ee
where $\epsilon \simeq .95$ was the value obtained from the simulations which we mentioned in the 
previous section.

From this false alarm probability, the second stage floating point operations coming only 
from false alarms is estimated to be, 
\be
\nop^{(2)} =  \bar{n}_c  \alpha \gamma  \left ( 6 \fsamp^{(2)} T  \log_2 (\fsamp^{(2)} T) \right ) ,
\label{eq:secondstagecost}
\ee
where $\gamma = \ntemp^{(2)} / \ntemp^{(1)}$ is the ratio of the number of templates in the 
second stage to the number of templates in the first stage, and $\alpha \gamma $ is the 
number of fine bank templates required to search the neighborhood of the trigger template 
that clicks. The Monte Carlo results indicate that $\alpha=2.5$ should be sufficient. We round this up to 
the integer value $3$.
Estimates of the computational cost and the speed-up factor over the flat search for 
these values of $\alpha$ are tabulated in Table \ref{tab:gain}. 
The computational cost estimates make the plausible assumption that the analysis will be dominated by the need to
dismiss false alarms, rather than processing true events. Therefore, the computational cost 
incurred when a signal is actually present in the data segment is not considered in these estimates. 
In our analysis, we have considered the 
second stage cost to be given by Eq. (\ref{eq:secondstagecost}). 

The online computational power for the two stages can be calculated by dividing the number 
of floating point operations $\nop^{(m)}, ~ m = 1,~2$ by the processed (zero-padded) length 
of the data segment $\tpad$:
\be
\nflop^{(m)} = \frac{ \nop^{(m)}}{ \tpad}.
\ee

The total online speed $\nflop^{\mathrm{(tot)}}$ is the sum of the two online speed for the two stages:
\be
\nflop^{\mathrm{(tot)}} = \nflop^{(1)} + \nflop^{(2)}.
\label{eq:floptotal}
\ee

\subsection{Choosing the optimal $\zeta_1$}
\label{sec:chooseZetaOne}

 We consider data segments of $512$ sec duration which 
accommodate chirps of maximum length $95\,\sec$ with sufficient padding 
corresponding to a minimum mass limit of $1.0 \,\msun$.
We consider first and second stage sampling frequencies of $512\,\Hz$ and $2048\,\Hz$ 
respectively. The cutoff frequency for the chirp in the first stage is set at 
$\fcut = 256\,\Hz$. 
Thus the number of data points to be processed are $2^{18}$ and $2^{20}$ for the coarse and 
fine banks respectively. The signals of minimal strength that can be observed, 
assuming a minimum detection probability of $95\%$ and $3\%$ mismatch between 
templates turns out to be $9.17$ \cite{monty1}. The first stage $\smin$ for the above 
mentioned cut-off can be calculated 
to be $92\%$ of 9.17 which is  $8.44$. If we choose a mismatch level of 
$\Gamma$ in order to tile the templates, the minimum $\sobs$ seen by each template 
will be $8.44\times\Gamma$. The first stage trigger level $\zeta_1$ must be chosen 
at approximately $8.44\times\Gamma-0.7$ so that the detection probability exceeds $95\%$ even in 
this stage. As discussed in \cite{monty1}, although minimum detection probabilities of $95\%$ are
chosen at each stage of the hierarchical search, the true detection probability is still $\sim 95\%$ because of the
strong correlation among the statistics of the first and second stages.

For relatively low values of $\Gamma$, the size of each individual rectangle is very large, so 
that fewer templates suffice to cover the parameter space. However, 
since $\zeta_1$ is also reduced in the process, there will be far too many 
crossings due to noise (false alarms) each of which
which need to be followed up by the fine bank {\it i.e} second stage filters. 
Thus the second stage cost increases.
On the other hand, when $\Gamma$ is high, the area covered by 
individual first stage tiles will 
be quite small, and therefore a larger number of tiles are needed to cover the parameter space thereby 
increasing the first stage cost.

\begin{figure}[h]
\centering
\includegraphics[width=0.55\textwidth]{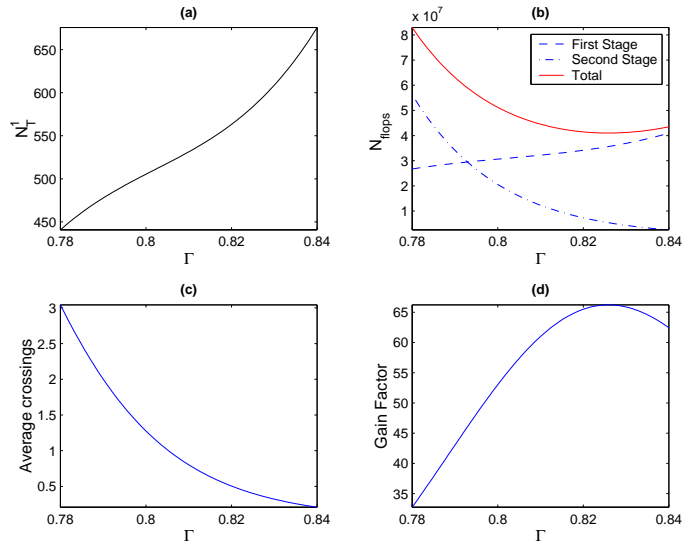}
\caption{(a) Number of trigger templates plotted against $\Gamma$. (b) Online cost 
of EHS algorithm plotted against $\Gamma$. (c) The average number of crossings in the first stage.
(d) The speed-up factor $\mathcal{G}$ factor defined as 
ratio of online flat search  cost to online total EHS cost is plotted against $\Gamma$. As the contour level $\Gamma$
is increased, the first stage cost increases because more trigger templates are required, while the second stage cost
goes down because the false alarm rate goes down. The total cost is minimum at an optimum value
$\Gamma_{\mathrm{opt}} \sim 0.8$ where the speed-up factor is maximum.}
\label{fig5}
\end{figure}

At some optimum value of $\Gamma = \Gamma_{\mathrm{opt}}$, the total cost reaches a minimum. This is the optimum contour level for laying the 
first stage templates. The optimum contour level also determines corresponding optimum 
value of the trigger threshold $\zeta_1$. From Fig \ref{fig5}, we find that $\Gamma_{\mathrm{opt}} \sim 0.8$.

\begin{table}[h]
\caption{Estimates of the computational cost for the two stage EHS algorithm and gain factors over the one
step flat search algorithm. $\Gamma$ is the
contour level of the ambiguity function used to lay the trigger tiles.
The speed-up factor ${\mathcal{G}}$ is defined as the ratio of required online computational cost for the
flat search to $\nflop^{(\mathrm{tot})}$. The online flat search cost is estimated to be $\sim$ 3.2 G-Flops.}
\begin{center}
\begin{tabular}{ll|cccccc}
			\hline
        &          
	& \ \ \ \ $\ntemp^{(1)}\ \ \ \ $ 
	& $\ \ \ \ \bar{n}_c\ \ \ \ \ $ 
	&$\nflop^{(1)}$  \ \ \ 
	& $\nflop^{(2)}$ \ \ \  
	& \ \ \ $\nflop^{\mathrm{(tot)}}$ \ \ \ \ 	
	& $\ \ \ \ \ \mathcal{G}\ \ \ \ 
						$\\ \hline \hline

$\alpha = 2.5$\ \ \ \ & $\Gamma = 0.81$ \ \ \ \ \ \ & 535 & 0.82 & 36.3 & 11.7  & 48.0 &   64 \\
                      & $\Gamma = 0.82$ \ \ \ \ \ \ & 564 & 0.51 & 38.3 & \ 6.8 & 45.1 &   68 \\
		    \hline
$\alpha = 3.0$\ \ \ \ & $\Gamma = 0.81$ \ \ \ \ \ \ & 535 & 0.82 & 36.3 & 14.0  & 50.3 &   61 \\
                      & $\Gamma = 0.82$ \ \ \ \ \ \ & 564 & 0.52 & 38.3 & \ 8.4 & 46.7 &   66 \\
		      \hline	
 
\end{tabular}
\label{tab:gain}
\end{center}
\end{table}

Using the above formulae, we are now in a position to estimate the online computational cost 
of the EHS. Recall that for the mass range considered, the longest chirp signal 
occurs for the smallest component masses $m_1 = m_2 = 1 ~ \msun$. For the lower cut-off of 
$f_l = 30$ Hz this signal is about 95 sec. Thus $\tpad \sim 512 - 95 = 417$ sec.
In the description of the flat search we mentioned that a total of $\ntemp = 11,050$ 
templates were required. Furthermore, implementing the trigger stage template
placement method described above, we find that a total of $\ntemp^{(1)} = 564$ templates would 
be required in this stage. Finally, setting $\zeta_1 = 6.05$ and using 
(\ref{eq:nc} -  \ref{eq:floptotal}) 
we have $\nflop^{(\mathrm{tot})} = 45.1$ M-Flops. This should be compared with the cost of a flat search 
given in Table \ref{tab:1}. For the similar mass limit, we find that the EHS algorithm 
reduces the online computational cost requirement over a flat search by a factor of 
$\sim 68$. The results for various choices of the parameters $\alpha$ and $\Gamma$ are 
summarised in Table \ref{tab:gain}.

\section{Conclusion}

We have investigated the performance of a new implementation of the two-step 
hierarchical search algorithm for detection of gravitational waves from compact inspiraling
binaries where the first stage is coarsely sampled in time, while retaining the earlier flavor
of hierarchy of templates over the mass parameters. The noise power spectral density used in
this work is that of the initial LIGO. We have described a new method of placing templates 
in the trigger stage of the hierarchical search. We have also clearly outlined methods to 
set up thresholds at the two levels based on probability arguments.

The calculation of false alarm probabilities and detection probabilities in this paper follow
from a basic assumption of Gaussianity and wide sense stationarity of the noise in the 
gravitational wave detectors. Also, we have assumed statistical independence of the output 
from the templates. However, we are aware that this may not be the case when the templates are
placed closely. We have performed Monte Carlo simulations to verify the effect of statistical
correlations between the templates. Our simulations show that there is very little correlation between 
the processed output of the first stage templates. We believe that this is because of the fact that
in this stage the template separations are quite large compared to earlier implementations of the hierarchical search.
We also find that even when the signal parameters are far away from the template centers, there is no appreciable loss
in detection probability. This result is however obtained for Gaussian noise. 
We estimate that an optimum implementation of the EHS algorithm would reduce the required online computing power by a
factor of $65-70$ over a flat search. The {\em{shape}} of the noise power spectral density has an important bearing in
the calculation of these numbers. 

If most of the compact binaries are likely to have equal companion 
masses, templates centered on the $\eta=1/4$ line will pick them up with a higher 
detection probability. But this entails using greater number of first stage templates 
since a significant fraction of these templates would spill over into the unphysical 
part of the parameter space - below the $\eta=1/4$ line. We have adopted a different 
strategy wherein we have tried to minimize this template spillage.
Our simulations  show
that even when the signal parameters lie close to the equal mass line, there is no 
appreciable loss in detection probability. However, to improve the probability of detection
in the scheme, templates can be centered on the bisector of the two boundary lines 
(a) $m_1=m_2 $ and (b) $m_1 = M_{\mathrm{min}}$. This strategy would increase the detection 
probability of the signals in the deemed parameter space without using more templates.
Note that by placing the tiles above the $\eta=1/4$ line, in the first stage, we are 
actually searching more than the mass range of $1.0 - 30.0 \ \msun$, because the tiles 
span a region which correspond to actual physical values of the individual masses but outside 
this range. If we include this region in our parameter space, then because the rectangles 
efficiently tile the 
total space - by definition - the gain factor could rise above 100.  

A search code for implementing the EHS algorithm is being written using the
LIGO Algorithm Library (LAL) \cite{lal} for the 
LIGO Data Analysis System (LDAS) \cite{ldas}.
It is to be noted that the existing LAL implementation of the template bank 
generation is based on the quadratic approximation of $\amb$. This is not 
suitable for the trigger stage of EHS which uses $\Gamma$ $(\simeq 0.82)$. 
At these values of $\Gamma$, the quadratic approximation to the ambiguity function   
is no more adequate - which means that it is not feasible to 
use the metric formalism \cite{owen1} to  place trigger stage templates 
in the  EHS scheme. 
Therefore, we plan to implement the approach of laying out the trigger stage templates using the 
algorithm described in this paper. The fine-bank templates 
will be layed out using the LAL bank package. 

For each event detected in the first layer of the hierarchy, the {\em approximate} time of arrival $t_a$
is already determined. Thus, only a smaller segment of the data at full sampling rate $\fsamp^{(2)}$ needs 
to be processed in the second stage. This could provide further savings in the second stage cost. 
However, as can be seen from Table \ref{tab:gain}, most of the computational cost is incurred in the trigger stage, so that the 
improvement in speed-up factor  will be small.

We have not addressed the issue of non-Gaussianity in this paper. However we believe
that tail features would lead to higher false alarm rates in the trigger stage - necessiating an 
upward revision of the $\Gamma_{\mathrm{opt}}$ from present values of $\sim 0.82$. This would lead 
to a reduction in the computational advantage of the EHS method as more templates would be needed at 
this stage. But perhaps putting another sieve in between the hierarchy layers may alleviate the 
problem.

\acknowledgments

The authors thank B.S. Sathyaprakash, S.D. Mohanty and P. Shawhan for useful 
discussions. A.S. Sengupta thanks LIGO Laboratory  for a three month visit to LIGO Caltech
and CSIR (India) for Senior Research Fellowship. S. Dhurandhar thanks Caltech for a visit during which 
writing of this paper was completed.   
The LIGO Laboratory works under cooperative agreement PHY-0107417. This work
was partially funded under NSF grant INT-0138459.
S. Dhurandhar and A. Lazzarini thank DST (India) and NSF (US) for the 
Indo-US collaborative programme. This document has been assigned LIGO Laboratory document
number LIGO-P020015-01.

\end{document}